\documentstyle[12pt]{article}

\catcode`\@=11

\newif\if@fewtab\@fewtabtrue

{\count255=\time\divide\count255 by 60
\xdef\hourmin{\number\count255}
\multiply\count255 by-60\advance\count255 by\time
\xdef\hourmin{\hourmin:\ifnum\count255<10 0\fi\the\count255}}
\def\draftdate{\number\day/\number\month/\number\year\ \ \ \hourmin }
\def\ps@draft{\let\@mkboth\@gobbletwo
    \def\@oddhead{}
    \def\@oddfoot
       {\hbox to 7 cm{$\scriptstyle Draft\ version:\ \draftdate$
       \hfil}\hskip -7cm\hfil\rm\thepage \hfil}
    \def\@evenhead{}\let\@evenfoot\@oddfoot}

\def\ceqno{\global\@fewtabfalse
    \ifcase\@eqcnt \def\@tempa{& & &}\or \def\@tempa{& &}
      \or \def\@tempa{&}
      \or\def\@tempa{}\fi\@tempa
{\rm(\theequation)}}

\def\aeqno#1{\global\@fewtabfalse
    \ifcase\@eqcnt \def\@tempa{& & &}\or \def\@tempa{& &}
      \or \def\@tempa{&}
      \or\def\@tempa{}\fi\@tempa
{\rm(\theequation#1)}}

\def\label#1{\ifnum\draftcontrol=1
 \global\def\draftnote{$\scriptstyle #1$}\fi
 \@bsphack\if@filesw {\let\thepage\relax
   \def\protect{\noexpand\noexpand\noexpand}%
\xdef\@gtempa{\write\@auxout{\string
      \newlabel{#1}{{\@currentlabel}{\thepage}}}}}\@gtempa
   \if@nobreak \ifvmode\nobreak\fi\fi\fi
  \@esphack}

\def\alabel#1#2{\label{#1}\global\@fewtabfalse
    \ifcase\@eqcnt \def\@tempa{& & &}\or \def\@tempa{& &}
      \or \def\@tempa{&}
      \or\def\@tempa{}\fi\@tempa
{\hbox to 3cm{\phantom{\rm(\theequation#2)}
\draftnote \hfil}\hskip -3cm {\rm(\theequation#2)}}}

\def\clabel#1{\label{#1}\global\@fewtabfalse
    \ifcase\@eqcnt \def\@tempa{& & &}\or \def\@tempa{& &}
      \or \def\@tempa{&}
      \or\def\@tempa{}\fi\@tempa
{\hbox to 3cm{\phantom{\rm(\theequation)}
\draftnote \hfil}\hskip -3cm{\rm(\theequation)}}}

\def\eqnarray{\def\draftnote{{}}\global\@fewtabtrue
\stepcounter{equation}\let\@currentlabel=\theequation
\global\@eqnswtrue
\global\@eqcnt\z@\tabskip\@centering\let\\=\@eqncr
$$\halign to \displaywidth\bgroup\@eqnsel\hskip\@centering\@eqcnt\z@
  $\displaystyle\tabskip\z@{##}$&\global\@eqcnt\@ne
  \hskip 1\arraycolsep \hfil$\displaystyle{##}$\hfil
  &\global\@eqcnt\tw@ \hskip 1\arraycolsep
$\displaystyle\tabskip\z@{##}$
\hfil  \tabskip\@centering&\global\@eqcnt\thr@@\llap{##}\tabskip\z@
\cr}

\def\endeqnarray{\@@eqncr\egroup
      \global\advance\c@equation\m@ne$$\global\@ignoretrue}

\def\@eqnnum{\hbox to 3cm{\phantom{\rm(\theequation)} \draftnote
                         \hfil}\hskip -3cm {\rm(\theequation)}}

\def\@@eqncr{\let\@tempa\relax
    \ifcase\@eqcnt \def\@tempa{& & &}\or \def\@tempa{& &}
      \or \def\@tempa{&}
      \or\def\@tempa{}
\fi\@tempa
\if@eqnsw
\if@fewtab\@eqnnum\fi
\stepcounter{equation}\fi\global
\@eqnswtrue\global\@eqcnt\z@\global\@fewtabtrue\cr}

\def\draftcite#1{\ifnum\draftcontrol=1#1\else{}\fi}

\def\@lbibitem[#1]#2{\item{}\hskip -3cm \hbox to 2cm
{\hfil$\scriptstyle\draftcite{#2}$}\hskip
1cm[\@biblabel{#1}]\if@filesw
     {\def\protect##1{\string ##1\space}\immediate
      \write\@auxout{\string\bibcite{#2}{#1}}}\fi\ignorespaces}

\def\@bibitem#1{\item\hskip -3cm \hbox to 2cm
{\hfil $\scriptstyle\draftcite{#1}$}\hskip 1cm
\if@filesw \immediate\write\@auxout
       {\string\bibcite{#1}{\the\value{\@listctr}}}\fi\ignorespaces}

\def\nsection#1{\section{#1}\setcounter{equation}{0}
                \def\theequation{{\thesection.\arabic{equation}}}}

\def\nappendix#1{\vskip 1cm\no{\bf Appendix #1}\def\thesection{#1}
                 \setcounter{equation}{0}
                 \def\theequation{{\thesection.\arabic{equation}}}}

\font\tendl=msbm10  scaled \magstep1
\font\sevendl=msbm7 scaled \magstep1
\font\fivedl=msbm5 scaled \magstep1
\font\tengl=eufm10  scaled \magstep1
\font\sevengl=eufm7 scaled \magstep1
\font\fivegl=eufm5 scaled \magstep1

\newfam\dlfam  
\textfont\dlfam=\tendl \scriptfont\dlfam=\sevendl
\scriptscriptfont\dlfam=\fivedl
\newfam\glfam  
\textfont\glfam=\tengl \scriptfont\glfam=\sevengl
\scriptscriptfont\glfam=\fivegl

\global\def\draftcontrol{0}
\catcode`\@=12

\def\pref#1{(\ref{#1})}

\newcommand{\be}{\begin{eqnarray}}
\newcommand{\en}{\end{eqnarray}\vs 0.5 cm}
\newcommand{\non}{\nonumber}
\newcommand{\no}{\noindent}
\newcommand{\vs}{\vskip}

\newcommand{\qq}{\begin{eqnarray}}

\newcommand{\qqq}{\end{eqnarray}}

\newcommand{\CO}{{\cal O}}

\newcommand{\CQ}{{\cal Q}}

\newcommand{\CZ}{{\cal Z}}

\def\goto{\rightarrow}

\def\det{{\rm det}}

\begin{document}
\def\PV{Pauli-Villars}
\def\FS{Faddeev-Slavnov}
\def\semi{;\hfil\break}
\title{On the Consistency of the Regularization of
Gauge Theories by High  Covariant Derivatives}
\author{\  Manuel Asorey and Fernando Falceto \\
\small Departamento de F\'{\i}sica
\small Te\'orica, Facultad de Ciencias\\
\small Universidad de Zaragoza
 E-50009 Zaragoza, Spain}
\date{  }
\maketitle
\vskip -8cm
\hfill\vbox to 8cm{
\hbox to 5cm{ \hfill
DFTUZ-95.3 \hskip-.5cm}
\vfill}
\vskip 1 cm

\begin{abstract}
        We show that regularization of gauge theories by higher
covariant derivatives and gauge invariant Pauli-Villars regulators
is a consistent method if the Pauli-Villars vector fields are
considered in a covariant $\alpha$--gauge with $\alpha\neq 0$
and a given auxiliary pre-regularization is introduced in order
to uniquely define the regularization. The limit $\alpha
\rightarrow 0$ in the regulating Pauli-Villars  fields is pathological
and the original Slavnov proposal in covariant Landau
gauge is not correct  because  of the appearance of
massless modes in the regulators which do not decouple when the
ultraviolet regulator is removed. In such a case the method does  not
correspond to the regularization of a pure gauge theory but that of a
gauge theory in interaction with  massless ghost  fields.
However, a minor modification of Slavnov method provides a
consistent regularization even for such a case.
The regularization that we introduce also solves the problem of
overlapping  divergences in a way similar to geometric
regularization and yields the standard values of the $\beta$
and $\gamma$ functions of the renormalization group equations.

  \overfullrule=0pt  \hyphenation{systems}
\vfill \noindent\baselineskip=12pt plus 2pt minus 1pt
\vfill\flushleft{{\bf PACS } numbers : 11.15.-q, 12.38.Bx}
 \thispagestyle{empty}
\end{abstract}

\vfill\eject
\nsection{Introduction}

In spite of the success of dimensional regularization
in perturbation theory, the existence of interesting non-perturbative
phenomena in gauge theories requires the introduction of
a non-perturbative regularization. Discretization of
space-time leads in a natural way to lattice regularizations which
preserve gauge invariance and have a non-perturbative
meaning. Unfortunately, the method does not seem to be appropriate
for the regularization of chiral,  supersymmetric or
topological  theories. The construction of a
non-perturbative gauge invariant regularization of gauge
theories in a continuum space-time has been a
challenging problem in gauge theories. A natural
candidate has always been a gauge invariant generalization of
Pauli-Villars methods involving high derivatives in the
action.

However, the regularization of gauge theories by high covariant
derivatives is plagued of difficulties. In scalar field theories the
addition of a number of derivatives to the kinetic term of the
action is enough to make the theory ultraviolet finite.
In gauge theories it is well known that the generalization of such
a method requires the introduction of covariant derivatives instead
of ordinary derivatives to preserve gauge invariance \cite{Slavnov0}.
Although
higher loop diagrams acquire by power counting a  negative
superficial  divergent dimension, the divergences of  one loop
diagrams are not smoothed by higher covariant derivatives
\cite{Slavnov0} \cite{Lee-Zinn}. One way of getting rid of the
remaining one loop  divergences could be the introduction
of an additional gauge invariant Pauli-Villars  regularization
\cite{Slavnov1} (see also \cite{Lee-Zinn}).
The concrete implementation of such regularization introduced
by Slavnov in Ref. \cite{Slavnov1} (see also Ref. \cite{Slavnov2}
for a review) encounters, however, two problems.
First, as it is well known
 the regularization does not remove all the
ultraviolet divergences. In fact, diagrams with external
Pauli-Villars lines  are not
regularized and their contributions to subdiagrams of higher loop
diagrams with external gluon lines is divergent \cite{AF1}
\cite{tesis}. The second problem has been  pointed out by a calculation
of  the $\beta$ and $\gamma$-functions of pure gauge theory
using the Slavnov regularization method \cite{mr}.
The result differs from the standard  values \cite{asym}
which are known to be   universal \cite{gross} \footnote{
There is a third problem which is associated to the ambiguity introduced
by the cancelation of  one loop divergences into the
definition of the regularization. Such an ambiguity,
which is a general feature of any
Pauli-Villars regularization, can be eliminated by a suitable choice
of an auxiliary pre-regularization prescription or by  a common
notation prescription for all one-loop diagrams involving Pauli-Villars
field propagators \cite{AFLL}}.

Any regularization method has to satisfy two main requirements.
First, it has to make
all Green functions of the theory finite. But this is not enough,
it has to satisfy one  further condition.
 The terms of the effective action that at a certain
order in perturbation theory are finite in the
original theory,  should  recover their exact finite values
after  the removal of  regulating mass parameters. In particular,
at a formal level, the regularized partition function
must  converge to the original one.

In this paper, we show that the method of regularization proposed in Ref.
\cite{Slavnov1} is not a regularization  of a pure
Yang-Mills theory in this sense but rather that of Yang-Mills
in interaction with a massless grassmannian ghost $\xi$
and a real commuting one $\phi$ in the adjoint representation,
\begin{eqnarray}
S(A)={1\over 4g^2}\int d^4x\, F^a_{\mu\nu}F_a^{\mu\nu}+
\int d^4x\, (D^\mu \bar\xi)_a (D_\mu \xi)^a
+{1\over 2}\int d^4x\, (D^\mu\phi)_a (D_\mu \phi)^a
\non\end{eqnarray}
The existence of
such a phenomenon is due to the fact that  the Pauli-Villars
counterterms introduced in Ref. \cite{Slavnov2} do not disappear when
the ultraviolet cutoff is removed. This fact does not
affect the one loop calculations
of the renormalization of the Chern-Simons coupling constant
using the Slavnov method, because the extra ghost interactions do not
generate unphysical pseudoscalar radiative corrections.

It is fairly easy, however, to
correct the form of the Pauli-Villars regulators to get a consistent
regularization. The problem of overlapping divergences requires
a more drastic modification of the standard prescription.

A solution of all these problems
was introduced in Refs. \cite{AF1}\cite{AF0} in terms of a
geometric interpretation of the regularization method. Other
consistent regularization methods with higher covariant derivatives
which are not affected by those problems were introduced in
Refs. \cite{Halpern}\cite{Warr}.

In this paper, we discuss a simpler higher covariant derivative
regularization method, closer to the original Slavnov proposal
and free of any of the difficulties mentioned above.
In section 2 we analyse the origin of the unphysical corrections
in the standard higher covariant
derivative regularization and we propose a method
to overcome the problem.  The advantages of the
regularization by Pauli-Villars field in $\alpha$-gauges are
investigated in section 3. In section 4 we
analyse  the problem of overlapping divergences and show how it
can be solved. In sections 5 and 6 we calculate the $\beta$ and
$\gamma$-functions of Yang-Mills theory at one loop
in the new regularization scheme and
compare the results with those obtained by geometric
regularization. Finally, in section 7 we present the
conclusions of
our work.

\vskip .5cm
\nsection{High Covariant Derivatives Regularization}

For simplicity we shall consider $SU(N)$ gauge theories,
although the results can be straightforwardly extended to arbitrary
gauge groups.
The euclidean action of   Yang-Mills
theory is given by
\begin{eqnarray}
S(A)={1\over 4g^2}\int d^4x\, F^a_{\mu\nu}F_a^{\mu\nu}
\non\end{eqnarray}
where $F^a_{\mu\nu}=\partial_\mu A^a_\nu -
\partial_\nu A^a_\mu+f^{abc} A^b_\mu A^c_\nu$  is the
field strength of the gauge field $A^a_\mu$.

The high covariant derivatives method  proposed by
Slavnov \cite{Slavnov1} (see \cite{Slavnov2} for a review)
proceeds by two steps.
The Yang-Mills action is replaced by its regularized version
\begin{eqnarray}
S_\Lambda(A)= {1\over 4g^2}
\int d^4x\,
F^a_{\mu\nu}[(I+{\Delta_\lambda /\Lambda^2})^n]
^{a'\mu\phantom{'}}_{a\phantom{'}\mu'} F_{a'}^{\mu'\nu},
\label{regacc}
\end{eqnarray}
where $\Delta_\lambda$ is the
covariant differential operator given by
$$(\Delta_\lambda) {}^{a\phantom{'}\mu\phantom{'} }_{a'\mu'}
=-D^2 {}^{a\phantom{'}}_{a'}
\delta^\mu_{\mu'}+2\lambda f^{a}_{a'c}F^{c\mu}_{\mu'}$$
in terms of the
covariant  derivative  $D_{\mu b}^a=\partial_\mu
\delta^a_b + f^a_{bc}A^c_\mu$ and $\lambda$ is an arbitrary real
constant.

The partition function for the regularized action in
$\alpha_0$--gauge is
\begin{eqnarray} \CZ=\int \prod_x dA(x)\, \det\,\partial^\mu
D_\mu \exp\left\{-S_\Lambda(A)-{1\over 2\alpha_0}
\int d^4x \,\partial^\mu A_\mu^a (I-\partial^2/\Lambda^2)^n
\partial^\nu A_\nu^a \right\}\non,
\end{eqnarray}
where we have  introduced the
operator  $(1-\partial^2/\Lambda^2)^n$  in the gauge fixing term
in  order to have an ultraviolet asymptotic behaviour
for the longitudinal modes of the propagator
similar to that of the transverse ones.
In this way, provided $n\geq 2$, all 1PI diagrams
with more than one loop acquire a negative degree
of divergence by  power
counting \cite{Slavnov0}. However, the degree of divergence of
one-loop 1PI diagrams is unchanged by the addition of
covariant derivatives. In other words, the theory is not completely
regularized by the simple fact of adding higher covariant
derivatives  to the action  as for the case of scalar
field theories \cite{Slavnov0}\cite{Lee-Zinn}.

Notice, however, that due to the regular behaviour of the gluonic
propagator the contributions in the effective action to
the ghost two point function
and gluon-ghost vertex are finite at one loop.
This implies that one loop divergences exclusively
arise  in diagrams with only external gluon lines,
and are given by
the following product of determinants
\begin{eqnarray}
\CZ_{\rm div}=\det(-\partial^\mu D_\mu)\, \det^{-1/2} \CQ
\label{oloop}
\end{eqnarray}
with
\begin{eqnarray}
\det^{-1/2} \CQ=\int \prod_{x}
d q(x)\,\exp&\Big\{&-{1\over2}\int d^4x\, d^4y\,
q_{\mu}^a (x){\delta^2 S_\Lambda\over
\delta A_\mu^a(x)\delta A_\nu^b(y)}
q_{\nu}^b (y)\cr\cr
&&-{1\over 2\alpha_0}
\int dx\, \partial^\mu q_\mu^a (I-\partial^2/\Lambda^2)^n
\partial^\nu q_\nu^a
\Big\}.
\non\end{eqnarray}
On the other hand, since  Faddeev-Popov ghost fields only
get finite renormalizations at one loop, the
divergences of $\CZ$ can be written in a gauge
invariant way.
This was the main observation made by Slavnov
\cite{Slavnov1} who proved that  one loop divergences of
Yang-Mills theory $\CZ_{\rm div}$     are
formally equal to those of
\begin{eqnarray}
\CZ_{\rm div}=\det (-D^2)\, \det^{-1/2} \CQ_0^L\label{oloopgi}
\end{eqnarray}
where
\begin{eqnarray}
\det^{-1/2} \CQ_0^L=\int \prod_{x}
dq(x)\,\delta(D^\mu q_\mu(x))\,\exp\left\{-{1\over 2}\int d^4x d^4y\,
q_{\mu}^a (x){\delta^2 S_\Lambda\over
\delta A_\mu^a(x)\delta A_\nu^b(y)}
q_{\nu}^b (y)\right\}.\label{CQ0}
\end{eqnarray}

We remark that all the determinants in \pref{oloopgi} are explicitly
gauge invariant. This fact can be understood as a consequence
of the absence of divergent radiative corrections to the
interaction of Faddeev-Popov ghost fields, which also
implies that the BRST symmetry is only renormalized
by  finite radiative corrections.
Moreover, gauge  invariance  is
not lost when we add   mass terms in \pref{oloopgi}.
Then, it seems natural to use these determinants as
the Pauli-Villars counterterms that subtract divergences
at one loop in a gauge invariant way. This is the Slavnov approach
introduced in Ref. \cite{Slavnov1} where
the author considers the \PV\ regulator
$$
\det^{-1/2} \CQ_{m}= \det(\Lambda^2m^2-D^2)\,
\det^{-1/2} \CQ_{m}^L
$$
with
\begin{eqnarray}
\det^{-1/2} \CQ_{m}^L= \int \prod_{x}
dq(x) &\delta(D^\mu q_\mu(x))\,\exp
\Big\{&- {1\over 2}\int d^4x d^4y\,
q_{\mu}^{a} (x){\delta^2 S_\Lambda\over
\delta A_\mu^a(x)\delta A_\nu^b(y)}
q_{\nu}^{b} (y)\cr\cr
&& - {1\over 2}m^2 \Lambda^2\int d^4x  \, q^2\Big\}.\label{CQM10}
\end{eqnarray}
The regularized partition function
\begin{eqnarray}
\CZ_{ \Lambda}&=&
\int\prod_x dA(x) \exp\left\{ -S_\Lambda(A)-{1\over 2\alpha_0}
\int d^4x \,\partial^\mu A_\mu^a (I-\partial^2/\Lambda^2)^n
\partial^\nu A_\nu^a
\right\} \cr\cr
&&\cdot\,\det\ (-\partial^\mu D_\mu)\prod_j
\det^{-{s_j/2}}\CQ_{m_j},\label{euno}
\end{eqnarray}
is, then, free of divergences at one loop provided
 the $s_j$ parameters are chosen so
that \begin{eqnarray}
&&\sum_j s_j+1=0.\label{PV}
\end{eqnarray}
Note that Pauli-Villars conditions do not involve the masses as
it is
usually the case. This is due to gauge invariance
and the high derivative
terms in the action that make finite the terms depending on $m$.
\vskip 3mm

Strictly speaking, in order to properly analyse the mechanism
of cancellation of one-loop divergences it becomes necessary to
introduce an auxiliary regulator to handle the different
divergences which appear by power counting \cite{Warr}\cite{AF1}
\cite{mr}.
There are several choices. Dimensional regularization
has the advantage of preserving gauge invariance
although it does not have
a non-perturbative interpretation.
If we  consider  such an  auxiliary regulator
the condition \pref{PV} is enough to make finite any one-loop
n-point function with external gluon lines (see appendix A for
an explicit computation). In section 4 we shall discuss other possible
choices of auxiliary regulator and the new conditions
we must impose in order to have a finite theory.

We have analyzed so far the main issues of the high covariant
derivatives regularization method proposed by  Slavnov
\cite{Slavnov1}. We shall now discuss the problems
raised by this method. In particular, we shall see that it is not a
suitable regularization of the theory.

Finiteness of Green functions is, of course, something
that any regularization must satisfy but this is not enough.
One further requirement
is that the terms of the effective action that at a certain
order in perturbation theory are finite in the
original theory, should converge to
the same value in the regularized
one when the regulator  $\Lambda$ is removed. In particular
at a formal level, the regularized partition function
$\CZ_{\rm reg}$
must  converge to the original one
$\CZ$  as the cut-off parameter $\Lambda$ goes to $\infty$.
This requirement is not satisfied by the regularization
presented in \cite{Slavnov1}.

The problem is that Pauli-Villars determinants $\det \CQ_m$ do not
converge
formally to a constant,
as they should, when the cutoff is removed. In fact, we have that
\begin{eqnarray}
\lim_{\Lambda\rightarrow \infty}\det^{-1/2} \CQ_{m}=
 \int \prod_{x}
d  q(x)\, \delta(D^\mu q_{\mu}(x))
\exp\left\{-{1\over 2}\int d^4x\,q^2(x)\right\}\label{Qlim}
\end{eqnarray}
that depends on $A$ through the delta functional
$\delta(D^\mu q_{\mu})$.
This can be seen more precisely if we
perform in \pref{Qlim} the change
of variables $q\rightarrow (q^\perp,\phi)$,
where $q_\mu=q_\mu^\perp+D_\mu \phi$,
\begin{eqnarray}
q_\mu^\perp=q_\mu-D_\mu D^{-2}D^\nu q_\nu\qquad {\rm and}
\qquad
 \phi =
 D^{-2}D^\nu q_\nu.
\label{chov}
\end{eqnarray}
The Jacobian  of the transformation is
$\det^{1/2}(-D^2)$. Therefore
\begin{eqnarray}
\lim_{\Lambda\rightarrow \infty}\det^{-1/2} \CQ_{m}=
 \int \prod_{x}d\phi(x)
&dq^\perp(x)&\delta(D^2\phi(x))\,
\exp\left\{-{1\over 2}\int d^4x (q^\perp-D\phi)^2(x)\right\}
\cr\cr
&&\cdot\,\det^{1/2}(-D^2),
\non\end{eqnarray}
and using the identity
$ \delta(D^2\phi)= \det^{-1} (-D^2)\,  \delta(\phi)$,
it reduces to
$$\lim_{\Lambda\rightarrow \infty}\det^{-1/2} \CQ_{m}=
\det^{-{1/2}}(-D^2),
$$
i.e. there is a net contribution of
the Pauli-Villars regulators to the effective action. This
fact explains why computations
of the $\beta$-function using this regularization \cite{mr}
do not agree
with the standard
value obtained by well established regularization methods
\cite{asym}. Using the Pauli-Villars condition \pref{PV}
one obtains that
the total anomalous contribution to the effective action
at one loop is given by
$$ \det^{{1/2}}(-D^2),$$
and it is  easy to compute its contribution to
the $\beta$-function
\begin{eqnarray}
\Delta\beta(g)=-{1\over 6}{g^3 N\over 16 \pi^2}.
\non\end{eqnarray}
If we add this contribution to the standard one
\begin{eqnarray}
\beta(g)=-{11\over 3}{g^3 N\over 16 \pi^2},
\non\end{eqnarray}
we
get the anomalous result
\begin{eqnarray}
 \beta^{\rm S}(g)=-{23\over 6}{g^3 N\over 16 \pi^2}
\non\end{eqnarray}
found in   Ref. \cite{mr}.
The same occurs for the $\gamma_A$ coefficient of
anomalous dimensions of the gauge fields.
If we add the extra contribution coming from
the determinant,
\begin{eqnarray}
\Delta\gamma_A(g)={1\over 6}{g^2 N\over 16 \pi^2},
\non\end{eqnarray}
to the standard value in $\alpha_0$ gauge
\begin{eqnarray}
\gamma_A(g)=\left({13\over 6}-{\alpha_0\over 2}\right){g^2 N\over 16
\pi^2},
\non\end{eqnarray}
we get  the value
\begin{eqnarray}
\gamma_A^{\rm S}(g)=\left({14\over 6}-{\alpha_0\over 2}
\right){g^2 N\over 16 \pi^2},
\non\end{eqnarray}
obtained in ref. \cite{mr}.
Furthermore, not only the divergent part of the effective action
picks up  unphysical contributions.
Five and higher point effective vertices, that
are already finite at one loop in the original theory, get
unphysical corrections which remain after  the cut-off
removal. Then the results differ from those obtained in the
original theory. The difference between both results does
coincide with the corresponding term  of
$ \log\det^{{1/2}}(-D^2)$.

The fact that
the specific prescription given by Faddeev-Slavnov
(with Pauli-Villars determinants computed in the covariant
Landau gauge)  is not correct does not mean, however, that
any method of regularizing gauge
theories by high covariant derivatives with  Pauli-Villars
is  necessarily inconsistent.
Once we have learned  the origin of the
problem, it is possible to implement a correct
regularization based on similar ideas.

It is in fact fairly easy to modify the prescription to
get a consistent regularization method. It is enough to
modify the Pauli-Villars contribution defining
the new regularized partition function by
\begin{eqnarray}
\CZ_\Lambda^{\rm new}&=
\int&\prod_x dA(x) \exp\left\{ -S_\Lambda(A)-{1\over 2\alpha_0}
\int d^4x \,\partial^\mu A_\mu^a (I-\partial^2/\Lambda^2)^n
\partial^\nu A_\nu^a
\right\} \cr\cr
&&
\det\ (-\partial^\mu D_\mu)
\prod_j\det^{{s_j/2}}(-{D^2})\,
\det^{-{s_j/2}}\CQ^L_{m_j}\,
\det^{s_j/2}(m_j^2\Lambda^2- D^2).
\label{Zrregg}
\end{eqnarray}

The partition function $\CZ_\Lambda^{\rm new}$ gives rise to finite
results for all n-point functions at one loop order
provided the Pauli-Villars condition \pref{PV} is satisfied.
Once again to properly define the regularized theory it is
necessary to introduce an auxiliary regulator. In appendix A we
show that within
dimensional regularization the condition \pref{PV} guarantees
the finiteness of the theory.
On the other hand the regularization  does not generate
any non physical correction in the ultraviolet limit and
maintains gauge invariance. It is a trivial exercise
to see that the computation of the radiative
corrections for the effective action at one loop,
using this regularization, gives rise to the correct
values for the $\beta$-function and anomalous dimensions
of the gauge field (see Section 5).

In this section we have pointed out the origin of the anomalous
result  for the beta function
and a way of overcoming the problem
with a minor change of the prescriptions of the
regularization. In order to gain a deeper understanding
of the regularization procedure we will examine, in next
section, the situation when we use covariant $\alpha$-gauge
instead of covariant Landau gauge to compute the Pauli-Villars
determinants. We shall find  that in such a case,
the straightforward implementation of the Faddeev-Slavnov 
prescription \cite{Slavnov1} provides a
consistent regularization of the theory.
In fact, the simple   addition of a mass term to all new
Pauli-Villars fields is enough to
define a   regularization which is
free of unphysical contributions.

\vskip .5cm
\nsection{Pauli-Villars regularization in $\alpha$-gauge}

Although the method of Pauli-Villars high covariant derivatives
regularization was formulated for an arbitrary $\alpha$--gauge
in the seminal Slavnov paper \cite{Slavnov1}, only the Landau
gauge ($\alpha=0$) case has been considered so far in the literature.
Let us analyse the general case in some detail.

It is easy to generalize the Slavnov arguments to show
that the  one loop divergences of the original Yang-Mills theory,
obtained from \pref{oloop} or \pref{oloopgi},
are the same that those of
\begin{eqnarray}
\det (-D^2)\, \det^{-1/2} \CQ_{0,\alpha} \,
\det^{1/2}(I-D^2/\Lambda^2)^n\label{oloopal}
\end{eqnarray}
where
\begin{eqnarray}
\det^{-1/2} \CQ_{0,\alpha}=\int \prod_{x}
dq(x)&\exp\Big\{&-{1\over 2}\int d^4x d^4y\,
q_{\mu}^a (x){\delta^2 S_\Lambda\over
\delta A_\mu^a(x)\delta A_\nu^b(y)}
q_{\nu}^b (y)\cr\cr
&&-{1\over 2\alpha}
\int dx D^\mu q_\mu^a (I-D^2/\Lambda^2)^n
D^\nu q_\nu^a
\Big\}.
\non\end{eqnarray}

To show this let us
perform the change
of variables $q\rightarrow(q^\perp,\phi)$ defined in \pref{chov}.
Taking into account the Jacobian of the transformation
$\det^{1/2}(-D^2)$,
we
obtain
\begin{eqnarray}
\det^{-1/2} \CQ_{0,\alpha}&=&\det^{1/2}
(-D^2)\clabel{CQ02}\cr&&\cdot\int \prod_{x}
dq^\perp(x) d\phi(x) \exp\Big\{-{1\over 2\alpha}
\int dx D^2 \phi^a(x) (I-D^2/\Lambda^2)^n
D^2\phi^a(x)\cr\cr
&&\quad-{1\over 2}\int d^4x d^4y
(q_{\mu}^{\perp a}+[D_\mu \phi]^a)(x){\delta^2 S_\Lambda\over
\delta A_\mu^a(x)\delta A_\nu^b(y)}
(q_{\nu}^{\perp b}+[D_\nu \phi]^b)(y)
\Big\}.
\end{eqnarray}
Now, since $S_\Lambda$ is gauge invariant,
\begin{eqnarray}
\int dx {\delta S_\Lambda\over
\delta A_\mu^a(x)}[D_\mu\phi]^a(x)=0,
\non\end{eqnarray}
and
\begin{eqnarray}
\int dx dy {\delta^2 S_\Lambda\over
\delta A_\mu^a(x)\delta A_\nu^b(y)}[D_\mu \phi]^a(x)q^b_\nu(y)
=-\int
dx  {\delta S_\Lambda\over
\delta A_\mu^a(x)}f^{a}_{bc} \phi^c(x) q^b_\nu(x).\label{gi}
\end{eqnarray}
Then, the vertex involving $\phi$ fields in the second term of the
exponent of \pref{CQ02} can be replaced by the right hand side
of \pref{gi}. This implies that all space-time derivatives
contained in such vertices act on the
external fields $A_\mu$ instead of the quantum fields $q^\perp$ and
$\phi$. This means that the corresponding diagrams are not divergent
if $n\geq 1$. Such a property is based on the same physical
reason that implied the finiteness of the one loop corrections
to the  ghost-gluon interactions.

Therefore, the divergent contribution of
$\det^{-1/2} \CQ_{0,\alpha}$ is given by
\begin{eqnarray}
&&\det^{-{1/2}}(D^2)\,
\det^{-{1/2}}(I-D^2/\Lambda^2)^n
\cr&&\cdot
\int \prod_{x}
dq^\perp(x) \exp\Big\{- {1\over 2}\int d^4x d^4y\,
q_{\mu}^{\perp a}(x){\delta^2 S_\Lambda\over
\delta A_\mu^a(x)\delta A_\nu^b(y)}
q_{\nu}^{\perp b}(y) \Big\}\cr\cr
&&=\det^{-1/2}\CQ_m^L\,
\det^{-1/2}(I-D^2/\Lambda^2)^n
\non\end{eqnarray}
and inserting the last expression for the divergences
of $\det^{-1/2} \CQ_{0,\alpha}$  into \pref{oloopal}
we obtain the expression \pref{oloopgi}.

Once we have shown that divergences of \pref{oloopal}
are equal to those of the original theory at one loop we can use
this last expression for the Pauli-Villars determinants.
Let us introduce
\begin{eqnarray}
\det^{-1/2} \CQ_{m,\alpha}=&\int & \prod_{x}
dq(x)\exp\Big\{- {1\over 2}\int d^4x d^4y\,
q_{\mu}^a (x){\delta^2 S_\Lambda\over
\delta A_\mu^a(x)\delta A_\nu^b(y)}
q_{\nu}^b (y)\clabel{CQal}
\cr\cr
&&-{1\over 2\alpha}
\int d^4 x \, D^\mu q_\mu^a (I-D^2/\Lambda^2)^n
D^\nu q_\nu^a - {1\over 2}m^2 \Lambda^2\int d^4x\,  q^2
\Big\}.\non\end{eqnarray}
One loop divergences can then be cancelled
by introducing a suitable
number of  massive
Pauli-Villars determinants. In fact, the
regularized theory can be defined from the partition function
\begin{eqnarray}
\CZ_\alpha&= \int &\prod_x  dA(x) \exp\left\{ -S_\Lambda(A)-
{1\over 2\alpha_0}
\int d^4x \,\partial^\mu A_\mu^a (I-\partial^2/\Lambda^2)^n
\partial^\mu A_\mu^a
\right\}\clabel{regal}
\cr\cr
&&\cdot\,\det\ (-\partial^\mu D_\mu) \prod_j\det^{-{s_j/2}} \CQ_{m_j,
\alpha_j}\,
\det^{s_j} (m_j^2\Lambda^2- D^2)\,
\det^{s_j/2} (I-D^2/\Lambda^2)^n.
\end{eqnarray}
{}From the previous analysis of divergences we conclude that
$\CZ_\alpha$ has, at one-loop, finite effective action
provided the $s_j$  are chosen
to satisfy the Pauli-Villars condition
$$\sum_j s_j +1=0.$$
In appendix A we verify that this condition it is enough
to cancel the $1/\epsilon$ divergences which appear in the different
one-loop diagrams by using dimensional regularization as auxiliary cut-off.
Notice, again, that due to the special
choice of the Pauli-Villars regulators
there are not constraints on $\alpha_0$, $\alpha_j$ and
$m_j$.
On the other hand BRST invariance at one-loop is
preserved because the new Pauli-Villars regulators
are explicitly gauge invariant.

Now it remains to check that
the determinants we added to the theory give a
contribution independent of the gauge field when the cutoff is removed.
If $\alpha\not=0$ we have
\begin{eqnarray}
\lim_{\Lambda \rightarrow \infty}
\det^{-1/2} \CQ_{m,\alpha}&= &
\lim_{\Lambda \rightarrow \infty}
\int \prod_{x}
dq(x)\exp\Big\{- {1\over 2 \Lambda^2}\int d^4x d^4y\,
q_{\mu}^a (x){\delta^2 S_\Lambda\over
\delta A_\mu^a(x)\delta A_\nu^b(y)}
q_{\nu}^b (y)\cr\cr
&&\qquad\quad
-{1\over 2 \alpha\Lambda^2}
\int d^4 x\, D^\mu q_\mu^a (I-D^2/\Lambda^2)^n
D^\nu q_\nu^a - {1\over 2} m^2 \int d^4x\,  q^2
\Big\}\cr\cr
&= &\int \prod_{x}
dq(x)\exp\Big\{- {1\over 2} m^2 \int d^4x\,  q^2
\Big\},\label{CQM7}
\end{eqnarray}
where  the previous equalities are to be understood up to
normalization constants.
We  obtain, then, the desired result since the
last expression is a constant independent of $A$.


We remark that in the limit in which the gauge fixing
parameters $\alpha_j$  vanish  we get a different theory.
Indeed, if we absorb $\alpha$  into the Pauli-Villars
 fields $q$, the determinant $\CQ_{m,\alpha}$ becomes
\begin{eqnarray}\non
\det^{-1/2} \CQ_{m,\alpha}=\int & \prod_{x}
dq(x)&\exp\Big\{- {\alpha\over 2}\int d^4x d^4y\,
q_{\mu}^a (x){\delta^2 S_\Lambda\over
\delta A_\mu^a(x)\delta A_\nu^b(y)}
q_{\nu}^b (y)\cr\cr
&&-{1\over 2}
\int d^4 x\, D^\mu q_\mu^a (I-D^2/\Lambda^2)^n
D^\nu q_\nu^a - {\alpha m^2 \Lambda^2\over 2}
\int d^4x\  q^2 \Big\}.
\end{eqnarray}
Therefore, if we take the limit $\alpha\rightarrow 0$
before removing the regulator $\Lambda$ we get
\begin{eqnarray}
\det^{-1/2} \CQ_{m,0}&=&\int \prod_{x}
dq(x)\,\exp\Big\{-{1\over 2}
\int d^4 x\, D^\mu q_\mu (I-D^2/\Lambda^2)^n
D^\nu q_\nu
\Big\}\cr\cr
&=&\det^{-1/2} (-D^2)\, \det^{-{1/2}}
(I-D^2/\Lambda^2)^n.\label{CQM9}
\end{eqnarray}

Now, although the second
factor in the last line of \pref{CQM9}
converges to the unity
in the limit ${\Lambda \rightarrow \infty}$,
 the first one gives an additional contribution
that indeed coincides with that
we already found in the previous section when we analyzed
the regularization in covariant Landau gauge.
Therefore, in this case we do no get a regularization of
pure Yang-Mills theory but a new theory which includes
a couple of scalar ghosts fields (one real bosonic, the other
complex grassmannian) interacting with the gauge field in the
adjoint representation.

The same fact can be seen
from a pure perturbative point of view.
Indeed, let us consider the propagator of the
Pauli-Villars fields,
$$
\delta^{ab}\left[{1\over p^2(1+p^2/\Lambda^2)^n+m^2\Lambda^2}
\left(\delta_{\mu\nu}-{p_\mu p_\nu\over p^2}\right)\,
+\,{\alpha\over p^2(1+p^2/\Lambda^2)^n+\alpha m^2\Lambda^2}
{p_\mu p_\nu\over p^2}\right].
$$
If $\alpha\not=0$ one loop diagrams
with a Pauli-Villars field running around the loop
and negative
degree of divergence, like five or higher point functions,
vanish in the limit $\Lambda\goto\infty$. Then, the new
determinants do not give
any contribution to the finite
parts of the theory.
However when we take the limit $\alpha\goto0$ before
removing the cutoff $\Lambda$ we see that
although the longitudinal part of the propagator, that behaves like
$${\alpha\over p^2(1+p^2/\Lambda^2)^n}
\cdot{p_\mu p_\nu\over p^2},
$$
seems to give a null contribution in this limit, this is not
true since there
are also vertices with a factor $\alpha^{-1}$,
coming from the gauge fixing term.
It is easy to show that divergences of these vertices
in $\alpha=0$ cancel out with the zeros
of the propagators and they contribute to a finite result
(independent of $m$) that does not vanish when we remove the cutoff.
It can be shown, again, that the new contributions are exactly those
of $\det^{1/2}(-D^2)$.

This singular character of Landau gauge was already noticed
in another context by Kennedy and King \cite{kk}. In our case
it means that standard Slavnov
regularization prescription  is correct in any
covariant $\alpha$-gauge  for the Pauli-Villars fields
except for Landau gauge  $\alpha=0$. In the latter case,
the prescription has to be
modified along the lines described in previous section.
\vskip .5cm
\nsection{Overlapping Divergences}

So far we have considered the regularization of the theory
at one loop. One could argue that, after adding high derivatives,
this is all we must consider because diagrams of more than one
loop are, then, finite by power counting and only one loop subdiagrams
diverge. This is true, but among those subdiagrams there are the
mixed loops of $A$ lines and $q$ lines that diverge and their
divergences are not cancelled by any other contribution.
This  problem, known as the overlapping divergence
problem, has remained for a long time as the main obstruction to
the implementation of the high covariant derivatives method for
non-abelian gauge theories as proposed in Ref. \cite{Slavnov2}.

There have been several
proposals to overcome this problem but not all of them
have been complete. In Ref. \cite{Slavnov3} it was
 suggested to use
different actions $S_{\Lambda,j}$ with
less covariant derivatives than
the original $S_\Lambda$ in  Pauli-Villars
determinants $\CQ_{m_j}^L$.
Explicitly, if we take
\begin{eqnarray}
S_{\Lambda,j}(A)= {1\over 4g^2}
\int d^4x
F^a_{\mu\nu}[(I+{\Delta_{\lambda_j} /\Lambda^2})^{n_j}]
{}^{a'\mu\phantom{'}
}_{a\phantom{'}\mu'} F_{a'}^{\mu'\nu}\qquad\hbox{with}
\qquad 2\leq n_j<n-1,
\label{lessder}\end{eqnarray}
and use $S_{\Lambda,j}$ instead of $S_\Lambda$ in \pref{CQM10}
and \pref{CQal} we see by power counting that the overlapping
divergences are absent. If in addition we
modify Faddeev-Slavnov prescription for covariant Landau
gauge as in Sec. 2, or use $\alpha$-gauge
for the Pauli-Villars determinants as in Sec. 3,
we can, at least in principle, define a theory
regularized at any order and free of unphysical
contributions. The problem, however, is
that in order to ensure finiteness of the theory
at one loop one has to change the Pauli-Villars condition \pref{PV},
and even the way to determine the new conditions
requires a more subtle analysis.

As mentioned before, the complete analysis
of the mechanism of cancellation
of one-loop divergences requires the
introduction of an auxiliary regulator to handle the different
divergences which appear by power counting
\cite{Warr}\cite{AF1}\cite{mr}.
The divergences of the different diagrams are
computed in the appendix by means of a dimensional regularization
as auxiliary regulator. The results show that the
 one loop  effective action
is finite if the condition
\begin{eqnarray}\label{DPV}
\sum_{j} s_j[\lambda_j^2(5n_j^2-18n_j+16)
- \lambda_j
(4n_j^2+10n_j+4)+4n_j^2+5n_j-7/3]=0
\end{eqnarray}
is satisfied. Here and below $n_0=n$ and  $s_0=1$ and we
assume $n_j\geq 2$.

Another  possible choice  of the auxiliary regularization
is to introduce a momentum
cut-off on one internal momentum in every loop \cite{Warr} or
still better, in order to have an unambiguous definition,
a momentum cut-off in {\it all} internal lines \cite{AF1}.
In general, this kind of auxiliary regularization
does not preserve gauge
invariance, but has the advantage of having a non perturbative
meaning \cite{AF1}. On the other hand as we shall discuss below
the breaking of gauge invariance can be controlled and it is possible
to tune the parameters of the regularization in order
to actually have a fully gauge invariant regularized theory.

If we use momentum cut-off as auxiliary regulator the cancellation
conditions become more stringent. In
Landau gauge all one-loop divergences
cancel out if the following conditions
are satisfied
\begin{eqnarray}
&\sum_j& s_j(n_j+1)^2+{1\over 6}s_0=0,\alabel{NPV}{a}\cr\cr
&\sum_{j\neq 0}& s_j=0,\aeqno{b}\cr\cr
&\sum_{j}& s_j[n_j+{2\over 3}]=0,\aeqno{c}\cr\cr
\sum_{j} s_j[\lambda_j^2(5n_j^2-18n_j+16)
-&\hskip-1mm\lambda_j&\hskip-2mm
(4n_j^2+10n_j+4)+4n_j^2+5n_j-7/3]=0.\quad\aeqno{d}\cr
\end{eqnarray}
Conditions (\ref{NPV}a) and (\ref{NPV}b) are necessary in order
to cancel linear divergences,  (\ref{NPV}c) and (\ref{NPV}d)
stand for the cancelation of
quadratic and logarithmic divergences, respectively.
The existence of linear divergences is a consequence of
our way of taking the momentum cut-off in {\it all} internal lines
and they may not appear for other choices of  auxiliary regulator.
Quadratic and linear divergences are absent for gauge invariant
pre-regularizations, while the logarithmic ones are
universal.
Note that
if all $n_j$'s and $\lambda_j$'s were
taken equal
the last two conditions would reduce to \pref{PV}.

Although the sharp
momentum cut-off regularizations break
gauge invariance, one could think that when the pre-regulator
is removed gauge invariance will be restored, because the
theory is finite. Unfortunately this
is wrong in general, and Slavnov-Taylor identities are violated
by  finite contributions generated by the momentum cutoff.
The anomalous contribution also appears as extra terms in
one-loop diagram identities of the diagrammatic approach
\cite{Holanda}.
Therefore one needs to impose
some additional constraints on the regulating parameters in order
to cancel such anomalous contributions to Slavnov-Taylor
identities \cite{AF2}. Indeed, although in principle
 gauge invariance is lost due to the auxiliary momentum cut-off
it is possible to adjust the free parameters
of the regularization in order to recover it.
Only a finite number of diagrams are
responsible for the breaking of the symmetry i.e
one loop diagrams with two, three
or four external legs. Thus, we only need  to compute the finite
parts of these diagrams and impose more constraints
on  the parameters of the regularization in order to ensure that the
terms that do not fulfil Ward identities cancell out.

The explicit calculation of these additional
conditions, though finite, is lengthy and
tedious and has not been carried out yet in
four-dimensional gauge theories. In three-dimensional space-times,
an explicit choice of parameters that preserve gauge invariance
of the regulated theory has been found in \cite{AF2}. Once all these
requirements are fulfilled we obtain a consistent gauge
invariant regularization which can be used for
cases where more conventional regularizations fail to give an
appropriate description of the  physical effects.

\vskip .5cm
\nsection{Renormalization Group Coefficients.}

In order to confirm the consistency of the regularization method
introduced in previous section, we  calculate here the one
loop renormalization group coefficients and we shall verify
that the results do coincide with  the universal values
for the $\beta$-function and the anomalous dimension of the gauge field.
These coefficients can be extracted from  the
divergent terms of the effective action in the limit   $\Lambda
\rightarrow \infty$.

We shall use the regularization scheme
introduced in section 4 with both gluons and
Pauli-Villars fields in $\alpha$ gauge,
i.e. we shall take expression
\pref{regal} with  $S_{\Lambda}$
replaced by $S_{\Lambda,j}$ in the computation
of $\det^{1/2}\CQ_{m_j,\alpha_j}$
and $n$ replaced by $n_j$ in $\det^{s_j/2} (I-D^2/\Lambda^2)^n$.

Taking  dimensional regularization as auxiliary regularization,
we obtain finite results for $D=4$ and  $\Lambda<\infty$, provided
\pref{DPV} is fulfilled.

Of course when the
ultraviolet regulator $\Lambda$ is removed divergences appear.
In this case we shall have only logarithmic divergences as these are
the only ones allowed by gauge invariance.
The computation
of the $\log\Lambda$ terms, which are relevant for
$\gamma$ and $\beta$ coefficients of the renormalization group,
can be considerably simplified by the following
observation.
Let us consider the corrections to the vacuum polarization tensor at
one loop. The only terms  that
generate a $\log\Lambda$ contribution are those
that were primitively ultraviolet divergent together with
those with a degree of divergence $-2$ in the infrared
(when the external momentum goes to zero). The former
were already computed to ensure finiteness of the theory and
the second ones get contributions from massless fields and
only from the part of their vertices with the
least number of derivatives.
The same is true for three points functions
changing the degree of divergence in the infrared to $-1$.
Below we present the different contributions
at one loop to gluon
and Faddeev-Popov ghost two
point functions as well as to the ghost-gluon-ghost
effective vertex.

The $\log\Lambda$ contribution to the vacuum polarization tensor
coming from loops of gluons in $\alpha_0$ gauge
(diagrams (1) and (2) of Figure 1) is given by:
\begin{eqnarray}
^g\Pi^{a_1 a_2}_{\mu_1 \mu_2}(q)&=&
	{g^2 N\over 16 \pi^2}\delta^{a_1
a_2}\left[4n^2+5n+2-\alpha_0 -\lambda(4n^2+10n+4)\right.\cr
&&\hskip 2cm\left.+ \lambda^2(5n^2-18n+16)\right](
q^2\delta_{\mu_1\mu_2}-
{q_{\mu_1}q_{\mu_2}})\log\Lambda+\CO(1)
\non\end{eqnarray}
for $n\geq 2$.

Loops of Faddeev-Popov ghosts do not generate any
$\Lambda$ dependence.
But there are
two different $\log\Lambda$ contributions
from loops of Pauli-Villars fields.
That coming from
$\det^{-1/2}\CQ_{m_j,\alpha_j}$ yields
\begin{eqnarray}
^{PV}\Pi^{a_1 a_2}_{\mu_1 \mu_2}(q)&=&
	{g^2 N\over 16 \pi^2}\delta^{a_1
a_2}\left[ 4n_j^2+16n_j/3-5/3- \lambda
(4n_j^2+10n_j+4)\right.\cr
&&\hskip 2cm\left.+ \lambda^2(5n_j^2-18n_j+16)\right](
q^2\delta_{\mu_1\mu_2}-
{q_{\mu_1}q_{\mu_2}}){\log\Lambda}+\CO(1),
\non\end{eqnarray}
where $n_j\geq 2$ is assumed.
The corresponding contribution from $\det^{1/2}(m^2\Lambda^2 -
D^2)^{n_j}$ reads
\begin{eqnarray}
^{PV_s}\Pi^{a_1 a_2}_{\mu_1 \mu_2}(q)=
	-{g^2 N\over 16 \pi^2}\delta^{a_1
a_2} {n_j\over 3}\left
(q^2\delta_{\mu_1\mu_2}-
{q_{\mu_1}q_{\mu_2}}\right ){\log\Lambda}+\CO(1).
\non\end{eqnarray}

Taking into account the finiteness condition \pref{DPV} the total
divergent logarithmic contribution of the vacuum polarization
tensor becomes
\begin{eqnarray}
\left({13\over 3}-{{\alpha_0}}\right){g^2 N\over 16
\pi^2}\delta^{a_1 a_2}\left
(\delta_{\mu_1\mu_2}q^2-
{q_{\mu_1}q_{\mu_2}}\right )\log{\Lambda}+\CO(1),\label{one}
\end{eqnarray}
and the renormalization group parameter associated to the
anomalous dimension  of the gauge field,
\begin{eqnarray}
\gamma_A(g)=
\left({13\over 6}-{{\alpha_0}\over 2}\right) {g^2 N\over 16
\pi^2}.
\label{anomA}
\end{eqnarray}

The ghost self-energy has  analogous  logarithmic divergent terms
\begin{eqnarray}
-\left({3\over 2}-{{\alpha_0}\over 2}\right){g^2 N\over 16
\pi^2}\delta^{a_1 a_2} q^2 \log{\Lambda}+\CO(1)
\label{two}
\end{eqnarray}
which contribute to
the renormalization of the anomalous dimension parameter of the
ghost field
$$
\gamma_c(g)= \left({3\over 4}-{{\alpha_0}\over
4}\right){g^2 N\over 16 \pi^2}.
$$

Finally, the relevant one-loop contribution to the
effective ghost-gluon-ghost vertex  obtained from the diagrams
(1) and (2) of figure 2 is given by
\begin{eqnarray}
-ig{{\alpha_0}}{g^2 N\over 16
\pi^2}f^{abc} p_\mu \log{\Lambda}+\CO(1).
\label{three}
\end{eqnarray}
The comparison of \pref{one}, \pref{two} and \pref{three} gives the
beta function  of the  coupling constant $g$,
\begin{eqnarray}
\beta(g)=-{11\over 3}{g^3 N\over 16
\pi^2}
\label{threes}
\end{eqnarray}
which agrees with the standard (universal) value obtained by other
methods \cite{asym}. Therefore, one
loop  results  for the $\beta$-function and the anomalous dimensions
parameters $\gamma_A$ and $\gamma_c$ confirm the validity of the
high covariant derivatives regularization method.

\vskip .5cm
 \nsection{Geometric Regularization.}

In this section we shall analyse, for the sake of completeness,
the results obtained by  geometric regularization method.
The method was introduced in
Refs. \cite{AF0}, \cite{AF1} and \cite{tesis},
in order to solve the problems of the conventional high
covariant derivative methods and to incorporate to the theory
geometric
elements of the space of gauge orbits which have a
non-perturbative meaning. One of the advantages of
the method is that it overcomes the Gribov problem giving a global
interpretation to the functional integral measure on the  gauge
orbit space.

The results listed below were first obtained in Ref. \cite{tesis} and the
reason to include them here is to show the
perturbative agreement of all consistent regularizations based on
the method of high covariant derivatives.

The calculation is also considerably simplified by the same
observation made in the previous section about the
connection of  logarithmic divergences on $\Lambda$ with
the infrared divergences on the one-loop corrections to the effective
action. The relevant contribution of gluon loops
to the vacuum polarization tensor $ \Pi^{ab}_{\mu\nu}$ is the same as
that obtained by
the method analysed in  the previous section \pref{one}.
However, the calculation of the ghost loops contribution is rather
different.

The central idea of geometric regularization is to
introduce two types of ghost fields: vector ghosts  associated
to the
metric of the orbit space
and scalar ghosts associated to the volume of the
fibres of gauge equivalent gauge fields.
Scalar ghosts have gauge invariant
interactions whereas vector ghosts do not.
The sum of their
contributions to the gluon effective action
does coincide with that of the
Faddeev-Popov determinant.
The splitting of the Faddeev-Popov determinant into the  product of
two determinants
\begin{eqnarray}
\det (-\partial^\mu D_\mu)=\det_{_L}^{1\over 2}
\left(\delta_{\mu}^{\nu}-D_\mu(D^2)^{-1}
D^\nu\right)\ \det^{1\over 2}
(-D^2) \end{eqnarray}
is based in the observation made
by Babelon and Viallet in Ref. \cite{Babelon}.
Here $\det_{_L}$ stands for the determinant
restricted to fields in the Landau gauge $\partial_\mu q^\mu=0$.
The geometric
interpretation of such a splitting allows to identify the
contribution of  vector ghosts with the functional riemannian volume
of the gauge orbit space, and that of the scalar ghosts with the
functional volume of the different gauge orbits.
The non gauge invariant character of the former, because
of the restriction of $\det_{_L}$ to a non invariant
subspace, can be understood as
a consequence of the necessary   choice of
coordinates to parametrize the space of gauge orbits.

The
global character of this geometric formulation of the
Faddeev-Popov gauge fixing method is preserved by geometric
regularization which replaces such a structure by a stronger
riemannian structure and two regularizing nuclear structures.
Because of the vectorial nature of those ghost fields it is natural
to impose on them the same cut-off that to the gauge fields and to
interpret it as the restriction to a submanifold of
the orbit space. This feature
provides a non-perturbative meaning to the
regularization method.

Now, since the relevant terms of the vertices are those which are
$\Lambda$ independent, the sum of the contributions generated by
nuclear, metric and scalar ghosts  reduces to that of
the  Faddeev-Popov ghosts. This follows from the
Babelon and Viallet splitting of
Faddeev-Popov determinant and from the theorem that establishes that
logarithmic divergences  are always preserved by determinant
factorizations (see \cite{AF1}).  Therefore, the global one
loop corrections to the anomalous dimension of the gauge field
$\gamma_A$ are also given by \pref{anomA} with $\alpha_0=0$.

The structure of geometric regularization implies that there  are
two different types of ghost self-energies. For vector (metric)
ghosts the logarithmic divergent contribution is given by
\begin{eqnarray}\non
{3\over 4} {g^2 N\over 16
\pi^2}\delta^{a_1 a_2} \left(\delta_{\mu_1 \mu_2}-{q_{\mu_1}
q_{\mu_2}\over q^2}\right)\log{\Lambda^2\over q^2}
\end{eqnarray}
which generates a
  renormalization of the anomalous dimension parameter of the
metric ghost field   $$\gamma_\psi= {3\over 4}{g^2 N\over 16
\pi^2}. $$

The remaining scalar ghost fields pick up a logarithmic divergent
contribution to its self-energy of the form
\begin{eqnarray}
{3}{g^2 N\over 16
\pi^2}\delta^{a_1 a_2} q^2 \log{\Lambda^2\over q^2}
\label{twooo}
\end{eqnarray}
which yields   a non-trivial anomalous dimension parameter
$$\gamma_\varphi= {3}{g^2 N\over 16 \pi^2}.$$

The renormalization of the coupling constant can be obtained from
the computation of  the  effective coupling of vector
or scalar ghosts to
the gauge field.
The one loop correction to the effective vertex of one gluon and two
vector ghosts, coming from  diagrams (1)-(5) of
figure 3, vanishes, while
the first perturbative correction to the effective
vertex with two scalar ghosts
is given by
\begin{eqnarray}
{9\over 4}{g^2 N\over 16
\pi^2}igf^{abc} (2p+q)_\mu \log{\Lambda^2\over q^2}.
\label{threee}
\end{eqnarray}

The value of the $\beta$-function obtained in both cases
\begin{eqnarray}
\beta=-{11\over 3}{g^3 N\over 16
\pi^2},
\label{threess}
\end{eqnarray}
and agrees with the standard
value obtained in previous sections.

\vskip .5cm
\nsection{Conclusions.}

The previous results show that the
high covariant regularization method is a consistent
regularization of gauge theories when one loop
divergences are properly regularized by means of
gauge invariant Pauli-Villars methods.

Because of the vectorial character of the gauge fields
the method requires the introduction of Pauli-Villars
vector fields.
If those vector fields are considered
in a generic $\alpha$-gauge the straightforward
generalization of
Slavnov method provides a consistent BRST
invariant regularization of the theory. However,
if those fields are considered in a covariant Landau gauge
($\alpha=0$),
the Slavnov method is not a proper regularization of the
theory. In such a case the method introduces  massless gauge fields
for each vector Pauli-Villars regulator and those fields do not
decouple from gauge particles in the limit when the ultraviolet
cut-off is removed. The existence of such a term is due to the
singular character of the limit $\alpha\to 0$ which does not
commute with the removal of the Pauli-Villars regularization mass
parameter. This fact explains the discrepancy found
in \cite{mr} in the calculation of $\beta$ and
$\gamma$-functions using the Slavnov regularization method.

The Pauli-Villars regularization introduced in this paper
does not suffer from any of those problems. It is a slight
modification of the original Slavnov proposal
which solves de problem of Landau gauge so that
the limit $\alpha\to 0$ is smooth. The calculation of
the $\beta$ and all $\gamma$-functions at one loop order
yields the standard universal values and
confirms the validity of the regularization method.

On the other hand we have shown how  it is
fairly easy  to overcome
the problem
of overlapping divergences
which is present in the Slavnov
proposal \cite{Slavnov2}.
The formal manipulations of the functional integral
which show the consistency of the regularization
can be substantiated by
introducing an auxiliary
regulator. Dimensional regularization provides a gauge
invariant pre-regulator which can be used to guarantee the consistency of
the theory. A  momentum
cut-off provides also a natural approximation which
has a meaning beyond perturbation theory,  but it
breaks the gauge symmetry.
Gauge invariance can be recovered once the spurious contributions
generated by the sharp momentum cut-off are cancelled
by tuning the parameters of
the regulators.
{}From a non-perturbative viewpoint
this regularization  method is useful in
cases where  the lattice
regularization is not suitable.

Indeed, the simple
fact that the problems discussed in this paper have
not been addressed earlier shows that the high covariant method
has been mainly used so far for formal arguments but
not explicit calculations. The only problem where the
method proved to be  useful was in the the regularization of
2+1 dimensional gauge theories with topological pseudoscalar
Chern-Simons term. However, in such a case the extra
unphysical contributions did not appear because the main
goal was to   calculate  the radiative
corrections to the Chern-Simons coupling constant, and
it is obvious that there are not anomalous unphysical
corrections at one loop to such an interaction, because of
the scalar character of the anomalous factor
$\det^{1/2}\ (-D^2)$. However, pathological contributions
will appear at  two loops and will provide a non-trivial
renormalization of the Chern-Simons coupling constant.

In summary, the high covariant derivative regularization method in
the formulation presented here is  fully consistent  and can be used
as an alternative regularization in those cases where other
methods fail either in perturbation theory or in the analysis
of non-perturbative effects.

\bigbreak\bigskip\bigskip\centerline {{\bf Acknowledgements}}
\nobreak
We thank  Benjamin  Grinstein for enlightening discussions.
One of us (M.A.) thanks
Prof. Gerard 't Hooft for correspondence.
We also acknowledge
to CICyT and E.U. (Human Capital and Mobility Program)
for partial financial support under grants AEN94-0218,
ERBCHRX-CT92-0035.

\nappendix{:}{\bf Cancellation of one-loop divergences}

\vskip 3mm
The special conditions which satisfy
the exponents and parameters of the different
regulators where chosen to ensure that  all one-loop divergent
contributions
which appear in the different regularized
partition function introduced in the paper,
cancel out to produce  finite effective actions.

In order to clarify the mechanism of cancellation we include
in this appendix the explicit calculations, using the dimensional
regularization as auxiliary regulator.
\vskip 3mm

	{\bf 1. One loop divergent contributions to the
          vacuum polarization tensor}

	{\it 1.1 Contribution of gluon loops.}

	In Landau gauge the divergent contribution of gluon loops
	 with regularized action $S_\Lambda(A)$   is given by
	 (diagrams (1) and (2) of Figure 1)
\begin{eqnarray*}
^g\Pi^{a_1 a_2}_{\mu_1 \mu_2}(q)&=
&-{1\over \epsilon}{g^2 N\over 16 \pi^2} \delta^{a_1
a_2}\Big[\Big( 4n^2+5n-7/3 -\lambda
(4n^2+10n+4)
\cr&&\hskip 3cm+ \lambda^2(5n^2-18n+16)\Big)(
q^2\delta_{\mu_1\mu_2}-
{q_{\mu_1}q_{\mu_2}})
\cr&&\hskip 3cm
+ {1\over6}q^2\delta_{\mu_1\mu_2}+ {1\over3}
{q_{\mu_1}q_{\mu_2}}\Big]+\CO(1)
\end{eqnarray*}
for $n\geq 2$. In the
case $n<2$ such a contribution cannot be obtained by taking  $n=1$
or $0$ in the previous expression. This fact can be easily
understood if we look
closely at the vertices of diagram
(1) in Figure 1. For large $n$ there is a particular term  in these
vertices  that generates divergences in the vacuum polarization
tensor and comes from taking as external line a gluon field from
$F_{\mu\nu}$, say the one on the right of \pref{regacc},
and as the two internal lines one from the $F_{\mu\nu}$ on the left
and the other from the most right Laplace-Beltrami operator.
Note that if we take the third $A$-field from
any other Laplace-Beltrami operator we do not
have a divergence in the two point function,
therefore the contribution  of this part of the vertex
to the $\epsilon$-divergent piece of $^g\Pi^{a_1 a_2}_{\mu_1 \mu_2}(q)$
does not depend on $n$
provided $n\geq 1$ but it is absent if $n=0$.
For $n=1$ there is a similar phenomenon
in the four point vertex.

The same results  hold for the theory in $\alpha_0$-gauge
\pref{euno}. In this case the extra $\alpha_0$
divergent contributions
arising in each diagram of Figure 1  cancel out in the
sum of both diagrams provided $n\geq 2$. Such a cancellation
does not occur for $n=0$.

{\it 1.2 Contribution of Faddeev-Popov ghost loops.}

	It is given by
\begin{eqnarray}\non
^{\phi \pi}\Pi^{a_1 a_2}_{\mu_1 \mu_2}(q) =
	{1\over \epsilon}{g^2 N\over 16 \pi^2}\delta^{a_1
a_2}\left
({1\over 6}q^2\delta_{\mu_1\mu_2}+ {1\over 3}
{q_{\mu_1}q_{\mu_2}}\right )+\CO(1)
\end{eqnarray}

{\it 1.3 Contribution of Pauli-Villars regulators.}

There are two different contributions. That generated
by $\det^{-1/2}\CQ^L_m$,
which is given by
\begin{eqnarray}
^{PV}\Pi^{a_1 a_2}_{\mu_1 \mu_2}(q)&=&
	- {1\over \epsilon}{g^2 N\over 16 \pi^2}\delta^{a_1
a_2}\left[ 4n^2+5n-5/3 +\lambda
(4n^2+10n+4) \right.\cr
&&\hskip 3cm-  \left.\lambda^2(5n^2-18n+16)\right]
(q^2\delta_{\mu_1\mu_2}-
{q_{\mu_1}q_{\mu_2}})+\CO(1),
\non\end{eqnarray}
for $n\geq 2$. And that induced by  $\det^{1/2}(m^2\Lambda^2 - D^2)
\det^{1/2}( - D^2)$, which reads
\begin{eqnarray}
^{PV_s}\Pi^{a_1 a_2}_{\mu_1 \mu_2}(q)=
{1\over \epsilon}{g^2 N\over 16 \pi^2}\delta^{a_1
a_2} {2\over 3}
(q^2\delta_{\mu_1\mu_2}-
{q_{\mu_1}q_{\mu_2}})
+\CO(1)\non
\end{eqnarray}
and is the same that that of $\det (m^2\Lambda^2 - D^2)$.

If we consider the Pauli-Villars fields associated to
$\det^{-1/2}\CQ_{m,\alpha}$ we get the following divergent contribution
\begin{eqnarray}
^{PV}\Pi^{a_1 a_2}_{\mu_1 \mu_2}(q)&=&
	-{1\over \epsilon}{g^2 N\over 16 \pi^2}\delta^{a_1
a_2}\left[ 4n^2+16n/3-5/3 -\lambda(4n^2+10n+4)\right.\cr
&&\hskip 3cm+\left. \lambda^2(5n^2-18n+16)\right](
q^2\delta_{\mu_1\mu_2}-
{q_{\mu_1}q_{\mu_2}})
+\CO(1),\non
\end{eqnarray}
for $n\geq2$. Note that, as it was expected from the
considerations of section 3, the $\alpha$-dependent
divergences generated by diagrams  (1) and (2) cancel out.

Finally, the contribution from $\det^{1/2}(m^2\Lambda^2 - D^2)^n$ is
\begin{eqnarray}
^{PV_s}\Pi^{a_1 a_2}_{\mu_1 \mu_2}(q)=
{1\over \epsilon}{g^2 N\over 16 \pi^2}\delta^{a_1
a_2} {n\over 3}\left
(q^2\delta_{\mu_1\mu_2}-
{q_{\mu_1}q_{\mu_2}}\right )
+\CO(1).\non
\end{eqnarray}
\vskip 3mm

{\bf 2. One loop divergent contributions to the three-point function}

	{\it 2.1 Contribution of gluon loops}

	In Landau gauge the divergent contribution of gluon loops
with regularized action $S_\Lambda(A)$  is given for $n\geq 2$ by
\begin{eqnarray}
^{g}\Gamma^{a_1 a_2 a_3}_{\mu_1 \mu_2 \mu_3}(q_1,q_2,q_3)&=&
{1\over \epsilon} {g^2 N\over 16 \pi^2}\left[ 4n^2+5n-9/4
+ \lambda^2(5n^2-18n+16)\right.\cr\cr
&&\hskip 2cm-\lambda(4n^2+10n+4) \left.\right] \Psi^{a_1 a_2 a_3
}_{{\mu_1 \mu_2 \mu_3}}
+\CO(1),\non
\end{eqnarray}
where
\begin{eqnarray}
\Psi^{a_1 a_2 a_3}_{{\mu_1 \mu_2 \mu_3}} =
i g f^{a_1 a_2 a_3}\left[ (q_1-q_2)_{\mu_3}\delta_{\mu_1 \mu_2}
+(q_2-q_3)_{\mu_1}\delta_{\mu_2 \mu_3}+ (q_3-q_1)_{\mu_2}
\delta_{\mu_1 \mu_3} \right].
\non\end{eqnarray}
for $n\geq 2$.
As before, the same results hold for the theory in
$\alpha_0$ gauge defined  in \pref{euno}.

{\it 2.2 Contribution of Faddeev-Popov ghost loops}

It is given by
\begin{eqnarray}
^{\phi \pi}\Gamma^{a_1 a_2 a_3}_{\mu_1 \mu_2 \mu_3}(q_1,q_2,q_3) =
-{1\over \epsilon}{g^2 N\over 16 \pi^2}{1\over 12} \Psi^{a_1 a_2 a_3
}_{{\mu_1 \mu_2 \mu_3}}
+\CO(1).\non
\end{eqnarray}

{\it 2.3 Contribution of Pauli-Villars regulators}

There are two different contributions. That induced by
to $\det^{-1/2}\CQ^L_m$
which is given by
\begin{eqnarray}
^{PV}\Gamma^{a_1 a_2 a_3}_{\mu_1 \mu_2 \mu_3}(q_1,q_2,q_3)&=&
{1\over \epsilon}{g^2 N\over 16 \pi^2}
\left[  4n^2+5n-5/3 -\lambda(4n^2+10n+4)\right.\cr
&&\hskip 2cm +\left.
\lambda^2(5n^2-18n+16)\right] \Psi^{a_1 a_2 a_3
}_{{\mu_1 \mu_2 \mu_3}}
+\CO(1),\non
 \end{eqnarray}
with $n\geq 2$, and that generated by  $\det^{1/2}(m^2\Lambda^2 - D^2)
\det^{1/2}( - D^2)$ which reads,
\begin{eqnarray}
^{PV_s}\Gamma^{a_1 a_2 a_3}_{\mu_1 \mu_2 \mu_3}(q_1,q_2,q_3)=
-{1\over \epsilon}{g^2 N\over 16 \pi^2} {2\over 3}\Psi^{a_1 a_2 a_3
}_{{\mu_1 \mu_2 \mu_3}}
+\CO(1),\non
\end{eqnarray}
and is the same that that of $\det (m^2\Lambda^2 - D^2)$.

If we consider the Pauli-Villars fields associated to
$\det^{-1/2}\CQ_{m,\alpha}$ we get the following divergent contribution
\begin{eqnarray}
^{PV}\Gamma^{a_1 a_2 a_3}_{\mu_1 \mu_2 \mu_3}(q_1,q_2,q_3)&=&
{1\over \epsilon}{g^2 N\over 16 \pi^2}\left[ 4n^2+16n/3-5/3-
\lambda(4n^2+10n+4)\right.\cr
&&\hskip 2cm\left. +
\lambda^2(5n^2-18n+16)\right]
\Psi^{a_1 a_2 a_3
}_{{\mu_1 \mu_2 \mu_3}}
+\CO(1),\non
\end{eqnarray}
for  $n\geq2$ and once again the $\alpha$-dependent
divergences generated by
the different diagrams
  cancel each other.

Finally, the contribution from $\det^{1/2}(m^2\Lambda^2 - D^2)^n$ is
\begin{eqnarray}
^{PV_s}\Gamma^{a_1 a_2 a_3}_{\mu_1 \mu_2 \mu_3}(q_1,q_2,q_3)=
-{1\over \epsilon}{g^2 N\over 16 \pi^2} {n\over 3} \Psi^{a_1 a_2 a_3
}_{{\mu_1 \mu_2 \mu_3}}
+\CO(1).\non
\end{eqnarray}
\vskip 3mm

{\bf 3. One loop divergent contributions to the four-point function}

	{\it 3.1 Contribution of gluon loops}

	In Landau gauge the divergent contribution of gluon loops
with regularized action $S_\Lambda(A)$  is given by
\begin{eqnarray}
^{g}\Gamma^{a_1 a_2 a_3 a_4}_{\mu_1 \mu_2 \mu_3
\mu_4}(q_1,q_2,q_3, q_4)&=&
{1\over \epsilon}{g^2 N\over 16\pi^2}
\Big[\left( 4n^2+5n-7/3 -\lambda
(4n^2+10n+4)
\right.\cr &&\hskip 1cm+
\lambda^2(5n^2-18n+16)\left.\right)
\Theta^{a_1 a_2 a_3 a_4}_{\mu_1 \mu_2 \mu_3 \mu_4}
+{1\over 6}
\Sigma^{a_1 a_2 a_3 a_4}_{\mu_1 \mu_2
\mu_3 \mu_4}\Big]
+\CO(1),\non
\end{eqnarray}
for $n\geq 2$,  where
\begin{eqnarray}\non
\Theta^{a_1 a_2 a_3 a_4}_{\mu_1 \mu_2 \mu_3 \mu_4}&=& -\big[
\phantom{+}f^{a_1 a_2 c}f^{a_3 a_4 c}(
\delta_{\mu_1 \mu_3}\delta_{\mu_2 \mu_4}-\delta_{\mu_1
\mu_4}\delta_{\mu_2 \mu_3})\cr
&&\ \ \ + f^{a_1 a_3 c}f^{a_2 a_4 c}(
\delta_{\mu_1 \mu_2}\delta_{\mu_3 \mu_4}-\delta_{\mu_1
\mu_4}\delta_{\mu_2 \mu_3})\cr
&&\ \ \ +f^{a_1 a_4 c}f^{a_3 a_2 c}(
\delta_{\mu_1 \mu_3}\delta_{\mu_2 \mu_4}-\delta_{\mu_1
\mu_2}\delta_{\mu_3 \mu_4})\big]
\end{eqnarray}
 and
\begin{eqnarray}\non
\Sigma^{a_1 a_2 a_3 a_4}_{\mu_1 \mu_2 \mu_3 \mu_4}&= &
{1\over N}
(
\delta_{\mu_1 \mu_2}\delta_{\mu_3 \mu_4}+
\delta_{\mu_1 \mu_3}\delta_{\mu_2 \mu_4}+
\delta_{\mu_1 \mu_4}\delta_{\mu_2 \mu_3})
\ (
f^{\alpha_1 a_1 \alpha_2}f^{\alpha_2 a_2 \alpha_3}
f^{\alpha_3 a_3 \alpha_4}f^{\alpha_4 a_4 \alpha_1}
\cr&&\hskip 6mm+
f^{\alpha_1 a_1 \alpha_2}f^{\alpha_2 a_2 \alpha_3}
f^{\alpha_3 a_4 \alpha_4}f^{\alpha_4 a_3 \alpha_1}
+
f^{\alpha_1 a_1 \alpha_2}f^{\alpha_2 a_3 \alpha_3}
f^{\alpha_3 a_2 \alpha_4}f^{\alpha_4 a_4 \alpha_1})
\end{eqnarray}
For $n \leq 1$ the contribution is different
for the reasons mentioned above.
As before, the same results hold for gluons in
$\alpha_0$ gauge.

{\it 3.2 Contribution of Faddeev-Popov ghost loops}

	It is given by
\begin{eqnarray}
^{\phi \pi}\Gamma^{a_1 a_2 a_3 a_4}_{\mu_1 \mu_2 \mu_3
\mu_4}(q_1,q_2,q_3, q_4) =
-{1\over \epsilon}{g^2 N\over 16\pi^2}
{1\over 6}
\Sigma^{a_1 a_2 a_3 a_4}_{\mu_1 \mu_2 \mu_3 \mu_4}
+\CO(1).\non
\end{eqnarray}

{\it 3.3 Contribution of Pauli-Villars regulators}

There are two different contributions. That generated
by $\det^{-1/2}\CQ^L_m$
which is given by
\begin{eqnarray}
^{PV}\Gamma^{a_1 a_2 a_3 a_4}_{\mu_1 \mu_2 \mu_3
\mu_4}(q_1,q_2,q_3, q_4)&=&
{1\over \epsilon}{g^2 N\over 16 \pi^2}\left[ 4n^2+5n-5/3
-\lambda(4n^2+10n+4)
\right. \cr &&\hskip 1.7cm\left.+\lambda^2(5n^2-18n+16) \right]
\Theta^{a_1 a_2 a_3 a_4}_{\mu_1 \mu_2 \mu_3 \mu_4}
+\CO(1),\non
\end{eqnarray}
where $n\geq 2$, and the contribution
form $\det^{1/2}(m^2\Lambda^2 - D^2)\det^{1/2}( - D^2)$
which reads,
\begin{eqnarray}
^{PV_s}\Gamma^{a_1 a_2 a_3 a_4}_{\mu_1 \mu_2 \mu_3
\mu_4}(q_1,q_2,q_3, q_4)=
-{1\over \epsilon}
{g^2 N\over 16 \pi^2} {2\over 3}
\Theta^{a_1 a_2 a_3 a_4}_{\mu_1
\mu_2 \mu_3 \mu_4}
+\CO(1),\non
\end{eqnarray}
and is the same that that of $\det (m^2\Lambda^2 - D^2)$.

If we consider the Pauli-Villars fields given by
$\det^{-1/2}\CQ_{m,\alpha}$ we get the following divergent contribution
\begin{eqnarray}
^{PV}\Gamma^{a_1 a_2 a_3 a_4}_{\mu_1 \mu_2 \mu_3
\mu_4}(q_1,q_2,q_3, q_4)&=&
{1\over \epsilon}{g^2 N\over 16 \pi^2}\left[ 4n^2+16n/3-5/3 -
\lambda(4n^2+10n+4)
\right.\cr &&\hskip 1.7cm+\left.
\lambda^2(5n^2-18n+16)\right]
\Theta^{a_1 a_2 a_3 a_4}_{\mu_1 \mu_2 \mu_3 \mu_4}
+\CO(1),\non
\end{eqnarray}
for $n\geq2$ and again the $\alpha$-dependent divergences generated by
the different diagrams cancel each other.

Finally, the contribution from $\det^{1/2}(m^2\Lambda^2 - D^2)^n$ is
\begin{eqnarray}
^{PV_s}\Gamma^{a_1 a_2 a_3 a_4}_{\mu_1 \mu_2 \mu_3
\mu_4}(q_1,q_2,q_3, q_4)=
	-{1\over \epsilon}{g^2 N\over 16 \pi^2} {n\over 3}
	\Theta^{a_1 a_2 a_3 a_4}_{\mu_1
\mu_2 \mu_3 \mu_4}
+\CO(1).\non
\end{eqnarray}

In this paper we have analyzed two regularization schemes.
The first one is that of sections 2 and 3
where we do not care about overlapping divergencies and the
number of high derivatives in gluons and Pauli-Villars fields
is the same. In such a case the cancellation of all $1/\epsilon$
divergence
follows from condition \pref{PV} which is enough to guarantee the
finiteness of the regularized theory in the limit
$\epsilon\rightarrow 0$.

In the second scenario, introduced in section 4 to cure the
overlapping divergences problem, the different Pauli-Villars
fields must have different number of derivatives and then the
condition for cancellation of $1/\epsilon$
 divergences  reduces to \pref{DPV}.

In conclusion, in both cases the cancellation of divergences   is
not only  a consequence of the formal identities derived from
functional integral methods, but it is a fact once
we introduce an appropriate auxiliary regulator.

\newpage
\section*{Figure Captions}
\bigskip

\def\fig{{\hfill\break \medskip \bf Figure}}
\noindent {\fig {\bf \ \ {1:}}}{ Radiative corrections to the vacuum
polarization involving gluon loops.}\hfill\break
\noindent {\fig {\bf $\,\,$ {2:}}}{ One loop contributions to the
3-vertex interaction of gluons with Faddeev-Popov
ghosts.}\hfill\break
\noindent {\fig {\bf \ \ {3:}}}{ One loop contributions to the
3-vertex interaction of gluons with scalar ghosts in
geometric regularization.}
\\

\end{document}